\documentclass[]{amsart}

\usepackage{lipsum}
\usepackage{amsmath}
\usepackage{amsfonts}
\usepackage{graphicx}
\usepackage{epstopdf}
\usepackage{algorithmic}
\usepackage{microtype}
\ifpdf
  \DeclareGraphicsExtensions{.eps,.pdf,.png,.jpg}
\else
  \DeclareGraphicsExtensions{.eps}
\fi
\usepackage{graphicx,epstopdf} 
\usepackage[caption=false]{subfig} 
\usepackage{physics}

\usepackage{url}
\usepackage{amsopn}

\providecommand{\keywords}[1]
{
  \small	
  \textbf{\textit{Keywords---}} #1
}

\makeatletter
\newcommand*{\addFileDependency}[1]{
  \typeout{(#1)}
  \@addtofilelist{#1}
  \IfFileExists{#1}{}{\typeout{No file #1.}}
}
\makeatother

\begin{document}

\author{Joseph V. Pusztay}
\thanks{Joseph V. Pusztay - University at Buffalo, Buffalo, NY. (josephpu@buffalo.edu)}

\author{Matthew G. Knepley}
\thanks{Matthew G. Knepley - University at Buffalo, Buffalo, NY. (kneply@buffalo.edu)}

\author{Mark F. Adams}
\thanks{Mark F. Adams - Lawrence Berkeley National Laboratory, Berkeley, CA. (mfadams@lbl.gov)}

\title{Conservative Projection Between Finite Element and Particle Bases}
\thanks{Submitted/accepted to SISC}
\thanks{JP was funded by the Department of Energy Exascale Computing Project WDMApp under subcontract S017663 and the Partnership for Edge Physics Simulation, Lawrence Berkeley National Laboratory subcontract 18083698. MGK was partially funded by the Department of Energy Applied Math Research under U.S. DOE Contract DE-AC02-06CH11357 and by the National Science Foundation under grant NSF CSSI: 1931524.}

\maketitle
\begin{abstract}
Particle-in-Cell (PIC) methods employ particle representations of unknown fields, but also employ continuum fields for other parts of the problem. Thus projection between particle and continuum bases is required. Moreover, we often need to enforce conservation constraints on this projection. We derive a mechanism for enforcement based on weak equality, and implement it in the PETSc libraries. Scalability is demonstrated to more than 1B particles.
\end{abstract}



\section{Introduction}
The Particle-in-Cell (PIC) method is a hybrid discretization, using a grid of finitely supported basis function to represent some fields, and a distribution of radial basis functions to represent other fields. The first discretization is usually called the \textit{mesh} basis, whereas the second is called the \textit{particle} basis. The particle basis can be useful for representing sharply localized quantities~\cite{BirdsallLangdon2018,hockney2021computer}, or fields in high dimension, or quantities that are more sensitive to artificial diffusion and dispersion. For example, it is common in plasma physics to represent clusters of electrons as a single ``macro particle'' eliminating the computational overhead of individual electron interactions~\cite{BirdsallLangdon2018}. In high dimension, particles are capable of more rapid convergence than mesh discretizations~\cite{AdamsBrennanKnepleyWang2021}. The self-consistent interaction of particles and fields in PIC methods can be important in resolving localized nonlinear effects~\cite{allanson2019particle}.

Particles may be represented using a shape function to capture particle width in the same way that finite element shape functions represent field behavior inside a mesh cell. Radial basis function methods are an example of this representation~\cite{fornberg2015primer}. In our experiments, we will use only the traditional delta function representation of particles, but the algebraic relations derived in Section~\ref{sec:math} remain valid for any shape function for which accurate quadrature is available. Each particle is assigned a weight, the coefficients of a basis function in the particle representation.

When employing PIC in large scale, long running simulations, particular attention must be paid to the choice of time marching integrator, selection of the mesh/particle bases, and projections between the bases to appropriately conserve mass, momentum, and energy. Conservation of these quantities has motivated a great deal of study in particle methods, particularly in the field of computational plasma physics \cite{doi:10.1063/1.4982054,CHEN20117018,MARKIDIS20117037,CHACON20131}. In this paper, we focus on a \textit{conservative} projection between basis representations, meaning projection which enforces the conservation of moments of the distribution. For example, conservation of the zeroth moment is equivalent to mass or charge conservation, the first moment to momentum conservation, and the second moment to energy conservation.

To begin, we will briefly define the PIC representation in \ref{sec:math}. Performance benchmarks for preconditioning and linear solves are discussed in \ref{sec:benchmarks}. We remark on the applications for these projection operators in \ref{sec:applications} before commenting on future improvements in \ref{sec:conclusion}

\section{Derivation}
\label{sec:math}
Suppose our system is comprised of small, massive bodies interacting with a potential. The standard representation is to assign a shape function to each, such that their weighted sum comprises the particle distribution function $f_P$. This distribution function can be interpreted as the probability of finding a particle at a given location $\vb{x}$ in configuration (or phase) space. Note that since we are working in phase space, $x$ would comprise both position and velocity. Thus, we can find by expected number of particles $n$ by integrating over all phase space
\begin{align}
  \int_\Omega f_P = n.
\end{align}
where $\Omega$ is the phase space domain. For simplicity, we use delta functions for this discussion, but our method does not depend sensitively on the choice of shape function. The full particle distribution function can then be written
\begin{align}\label{eq:particleField}
  f_P &= \sum_p w_p \delta(\vb{x} - \vb{x_p})
\end{align}
where $\vb{x}$ represents the configuration space variable, $\vb{x_p}$ is the particle location and velocity, and $w_p$ the particle weight. The finite element representation, using function space $\mathcal{V}$, is given by the weighted sum of basis function,
\begin{align}
  f_{FE} = \sum_i f_i \phi_i(\vb{x})    
\end{align}
where $\phi_i \in \mathcal{V}$ denotes the basis function, and $f_i$ the associated coefficient.

At this point, we must determine what we mean when we say these two expressions represent the same field. For instance, in phase space, the two systems need to contain the same mass, represented by equivalence of the integrals
\begin{align}
  \label{eq:mom0}
  \int_\Omega f_{FE} = \int_\Omega f_P.
\end{align}
In PIC codes, it is very common to enforce a tensor produce structure on the phase space. For example, many collision operators, such as the Landau operator, are completely localized in space. This converts the integral over phase space into an integral over velocity space $\Omega_V$~\cite{AdamsHirvijoki2017,mollen_adams_knepley_hager_chang_2021}. This means that equality of the first moment in phase space,
\begin{align}
  \label{eq:mom1}
  \int_\Omega \vb{x} f_{FE} = \int_\Omega \vb{x} f_P.
\end{align}
becomes the conservation of momentum
\begin{align}
  \int_{\Omega_V} m \vb{v} f_{FE} = \int_{\Omega_V} m \vb{v} f_P 
\end{align}
when multiplied by the mass $m$. In the same way, conservation of the second moment
\begin{align}
    \label{eq:mom2}
    \int_\Omega |\vb{x}|^2 f_{FE} = \int_\Omega |\vb{x}|^2 f_P
\end{align}
becomes the conservation of kinetic energy
\begin{align}
  \int_{\Omega_V} \frac{1}{2} m |\vb{v}|^2 f_{FE} = \int_{\Omega_V} \frac{1}{2} m |\vb{v}|^2 f_P.
\end{align}

These conditions are all specific instances of what we will call \textit{weak equivalence}, namely the equivalence of two expressions in the subspace spanned by some set of functions,
\begin{align}
    \label{eq:weakeq}
    \int_\Omega \phi_i\, f_{FE} = \int_\Omega \phi_i\, f_P \quad \forall \phi_i \in \mathcal{V} 
\end{align}
The finite dimensional analogue of \ref{eq:weakeq} is perhaps easier to interpret. We begin by expanding each field into its components~\cite{kirby2004,KnepleyBrownRuppSmith13}
\begin{align}
  \label{eq:testFunction}
  \int_\Omega \phi_i \sum_j f_j \phi_j &= \int_\Omega \phi_i \sum_p  w_p \delta(\vb{x} - \vb{x_p}), \\
  \sum_j f_j \int_\Omega \phi_i \phi_j &= \sum_p  w_p \int_\Omega \phi_i \delta(\vb{x} - \vb{x_p}).
\end{align}
We can rewrite this using linear algebraic notation,
\begin{align}\label{eqn:conservativeProjection}
  M f = V w
\end{align}
where $M$ is the FEM mass matrix
\begin{align}
    M = \int_\Omega \phi_i \phi_j,
\end{align}
$V$ the particle mass matrix
\begin{align}
    V = \int_\Omega \phi_i \delta(\vb{x} - \vb{x_p}),
\end{align}
$f$ the vector of finite element coefficients, and $w$ the vector of particle weights. The particle mass matrix entries are the evaluation of FEM basis functions at the particle positions, with the basis function determining the row and the particle determining the column. This simple equation allows us to convert particle weights to finite element coefficients, and vice versa, while preserving all the moments which are contained in the finite element space. Note, that as particle positions or momenta change, $V$ will likewise change.

Computation of $M$ and $V$ is straightforward when delta functions are used to express $f_P$. If more general shape functions are used, accurate calculation of the elements of $V$ could become expensive since accurate quadrature for the overlap of the FEM basis functions and the particle shape function would be necessary.

Inversion of the finite element mass matrix, M, is straightforward given the favorable conditioning~\cite{Wathen1987}, and we employ conjugate gradient method preconditioned with ILU(0) in~\ref{sec:benchmarks}. The particle mass matrix need not be square, and thus a solver for over/underdetermined systems is necessary~\cite{PaigeSaunders1982}, in most cases with preconditioning for scalability. Typical preconditioning uses an approximation to the normal equations, so we recall that
\begin{align}
  V_{ip} = \int_\Omega \phi_i \delta(x - x_p),
\end{align}
so that our normal equations are given by
\begin{align}
  V^T V &= V_{qi} V_{ip}, \\
        &= \left(\int_\Omega \phi_i \delta(x' - x_q)\right) \left(\int_\Omega \phi_i \delta(x - x_p)\right), \\
        &= \phi_i(x_q) \phi_i(x_p)
\end{align}
In our experiments, preconditioning with the normal equations is quite effective, and reduces the number of iterations by one to two orders of magnitude.

\section{Numerical Results}
\label{sec:benchmarks}

We have implemented the projection algorithm from Section~\ref{sec:math} using the PETSc libraries~\cite{petsc-user-ref,petsc-efficient,petsc-web-page}. Our particle basis is embodied by a DMSwarm object~\cite{2017EGUGA..1910133M}, and the finite element mesh and function space by a DMPlex object~\cite{KnepleyKarpeev09,LangeMitchellKnepleyGorman2015}. The PETSc suite of linear solvers is used to solve the algebraic equations for conservative projection, Eq~\ref{eqn:conservativeProjection}.

The principal application for this algorithm is currently the projection of a particle field to a finite element space for the purpose of applying a collision operator, in particular the Landau operator~\cite{AdamsBrennanKnepleyWang2021,mollen_adams_knepley_hager_chang_2021}. Therefore, we will restrict our phase space to the velocity component. We produce a mesh covering this space that defines our finite element space, and a set of particles which define the particle field. We project the particle field to the finite element space and back, checking for conservation of moments at each step.

Runtime of the projections solves is controlled principally by the choice of preconditioner. A strong preconditioner is also necessary in order to reach high accuracy in the solve. We require errors on the order $10^{-14}$ in our tests. Inversion of the finite element mass matrix in the particle deposition step is well understood, and our choice of CG/ILU(0) is sufficient. However, the weakly coupled, block nature of the particle mass matrix presents an opportunity to study block preconditioners for the normal equations, using LSQR~\cite{PaigeSaunders1982} as the Krylov solver. To this end, a \textit{weak scaling} study is presented for preconditioned LSQR preconditioned with Block Jacobi on the normal equations using sparse LU on each block, as well as Additive Schwarz (ASM) with ICC(0) on each block. All tests were performed using standard PETSc examples, making it straightforward for a reader to reproduce the results, and to alter the solvers, such as replacing our choice of Krylov solver or block subsolver. The numerical results present a study on fixed particle number per cell, with an increasing number of cells such that the number of cells across a fixed domain remain fixed on each node.

\begin{figure}[tbhp]
    \label{fig:grid}
    \centering
    \subfloat[FE Coefficients]{\label{fig:fe_coef}\includegraphics[width=.5\textwidth]{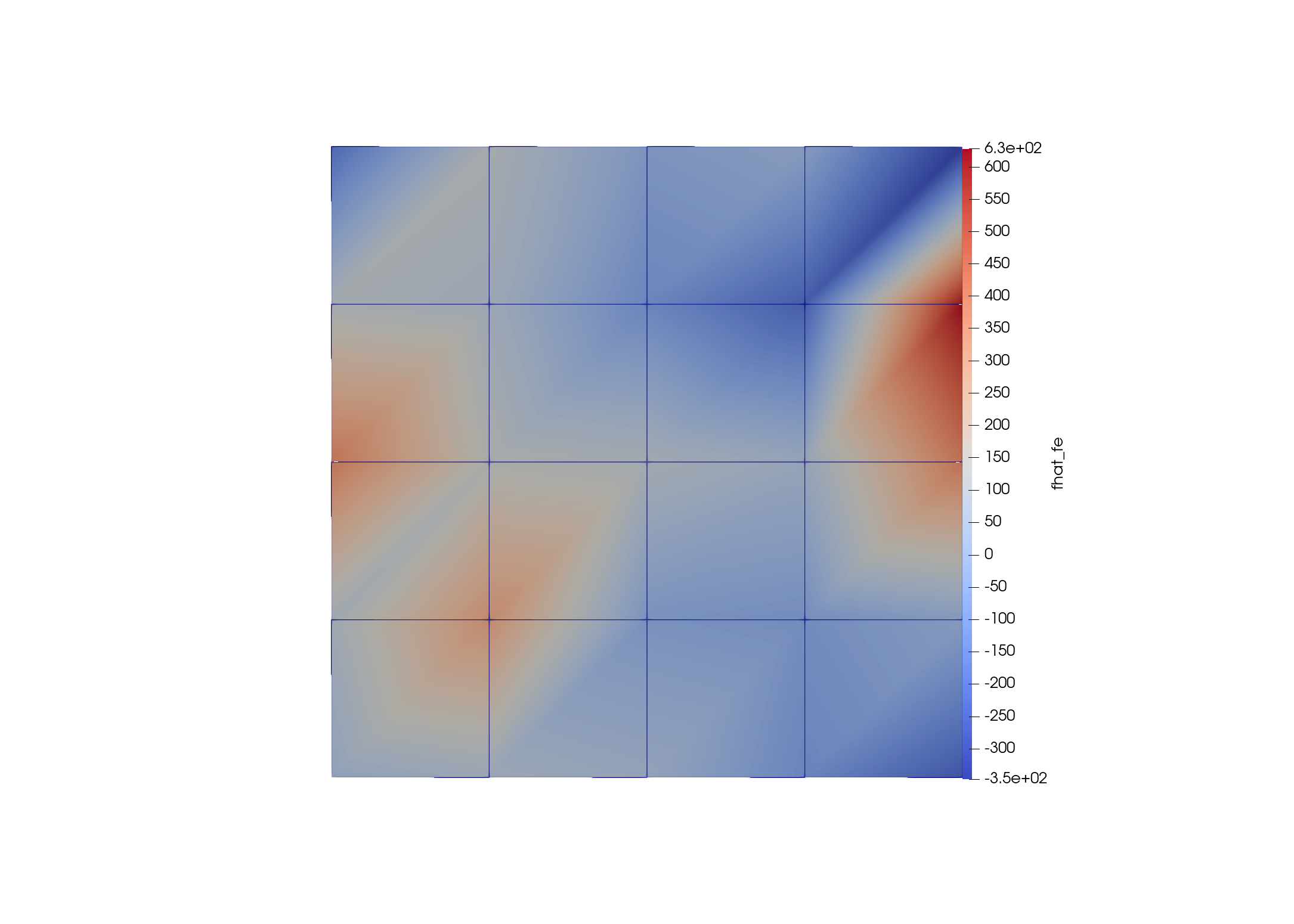}}
    \subfloat[Particles]{\label{fig:particles}\includegraphics[width=.5\textwidth]{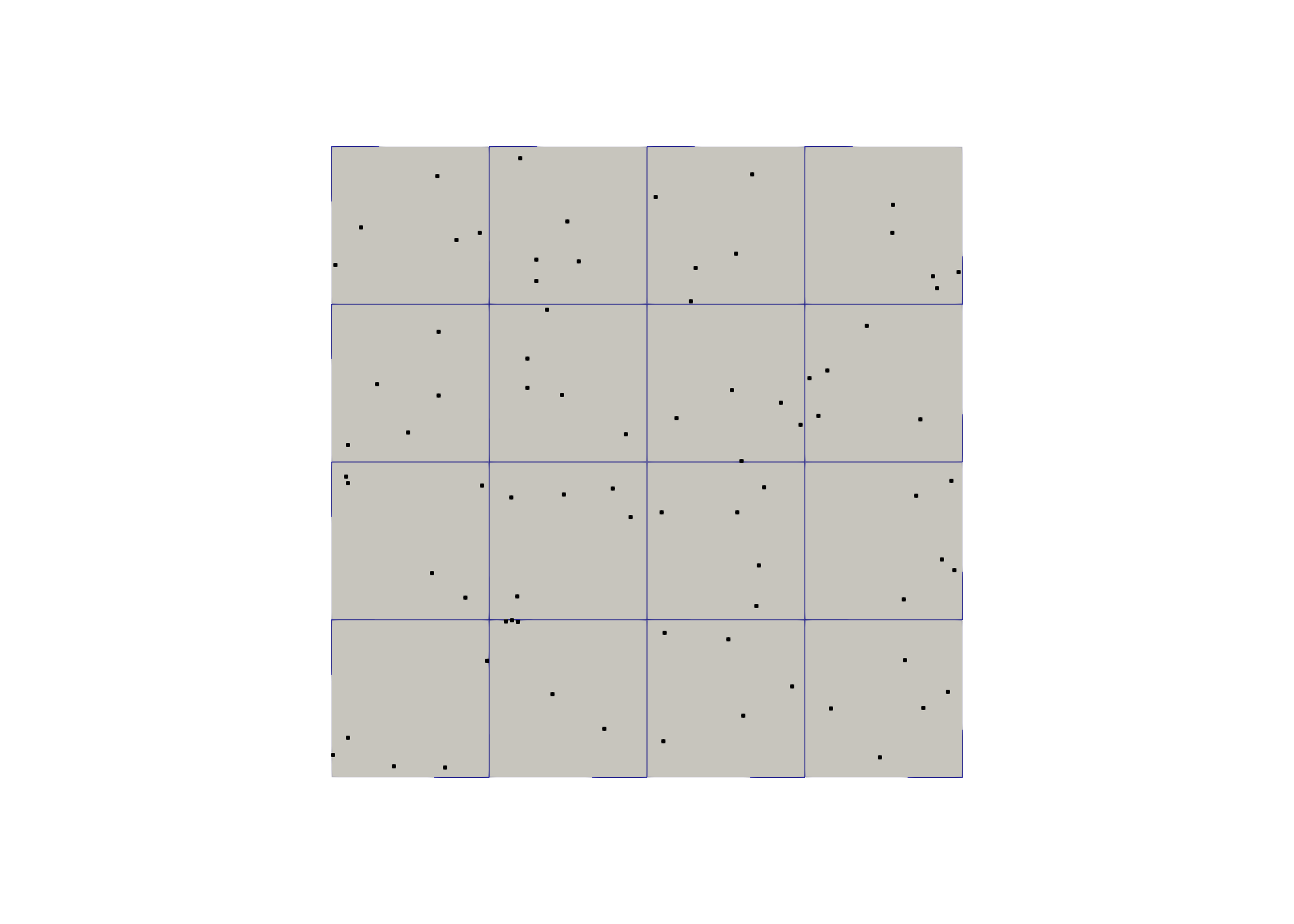}}\\
    \caption{Visualization of the finite element field (a) arising for a distribution of particles (b) on a 4x4 grid with 5 particles per cell.}
    \label{fig:scaling}
\end{figure}

All performance studies were conducted on the Geosolver cluster, located at the University at Buffalo Center for Computational Research. The Geosolver consists of 100 Intel Xeon Gold 2.6GHz 12-core 6126 processors, with 8GB of RAM per core, for a total of 1200 cores across 50 nodes with Intel Omnipath interconnects. We present the PETSc log for a representative run consisting of $>1$ billion particles in the supplementary materials associated with this article, as well as instructions for solver configurations in PETSc for duplication of the presented results. We present  \ref{fig:grid} as a smaller version of the system being solved in both a field and particle representation.

Weak scaling is demonstrated with an initial configuration of $256 \times 256 = 65,536$ quadrilateral, or tensor, cells with 200 particles per cell, and a simplicial mesh of $128 \times 128 = 16,348$ cells with 200 particles per cell. Particle coordinates within each cell are randomized for each run. One node using 24 MPI processes was chosen as a starting point. As the number of nodes were doubled, the domain was further subdivided by increasing the cell count by a factor of two, resulting in each node holding a $256 \times 256$ cell mesh in the tensor case, or $128 \times 128$ in the simplicial case. All tests use a quadratic element space, with third order quadrature on the cells, however the simulation can change the order and quadrature with command line arguments.

\begin{figure}[tbhp]
    \centering
    \subfloat[Tensor Cells]{\label{fig:ptof_error}\includegraphics[width=.5\textwidth]{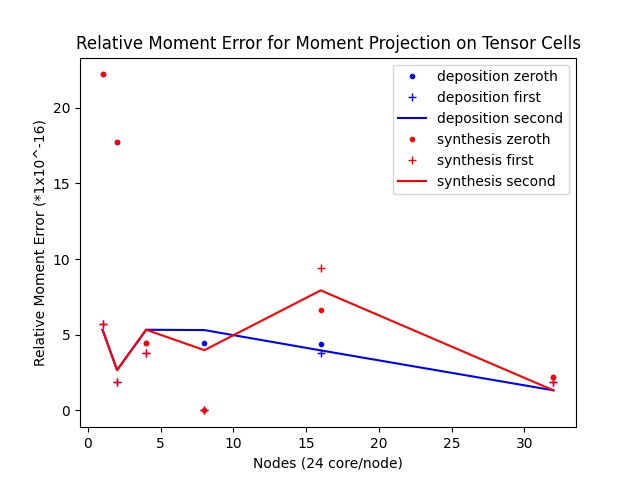}}
    \subfloat[Simplices]{\label{fig:ftop_error}\includegraphics[width=.5\textwidth]{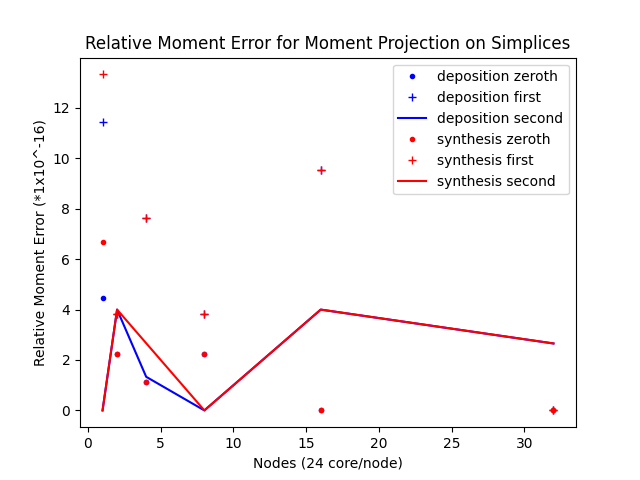}}\\
    \caption{Moment error for the zeroth (mass), first (momentum), and second (energy) moments after projection from particles to FE (deposition) and FE to particles (synthesis) in a single particle per cell case for simplices and tensor cells. The particle is located at the centroid of each cell, and demonstrates a single particle is sufficient to reconstruct the field with relative moment error within $100*\epsilon_{machine}$~\cite{MachineEpsilon} regardless of overall mesh size.}
    \label{fig:1ppc_error}
\end{figure}

Before we discuss the scaling with respect to a moderate, fixed particle number per cell, it is useful to observe the capabilities of these projections on a smaller scale. To do so, we configure an initial 256x256 tensor mesh on one node, with each cell containing one particle positioned at its centroid. Particle deposition, the projection of the particle field into the FE space (ptof), and particle synthesis, projection from the FE to particle space (ftop), are then performed, and the relative error between the moments of the bases are observed. \ref{fig:1ppc_error} plots the relative error of each moment as a function of mesh size, and demonstrates the ability to accurately reconstruct the field in the particle synthesis step with a single particle. An additional weak scaling study is presented with a "checkerboard" configuration in the respective subsections for each solver configuration. The particle numbers in the checkerboard configuration alternate between five and six particles for each given cell of the mesh.

\subsection{Block Jacobi+LU Weak Scaling}
\label{subsub:bj_wscaling}
\begin{figure}[tbhp] 
    \centering
    \subfloat[256x256 tensor cells per node]{\label{fig:weak_scaling}\includegraphics[width=.45\textwidth]{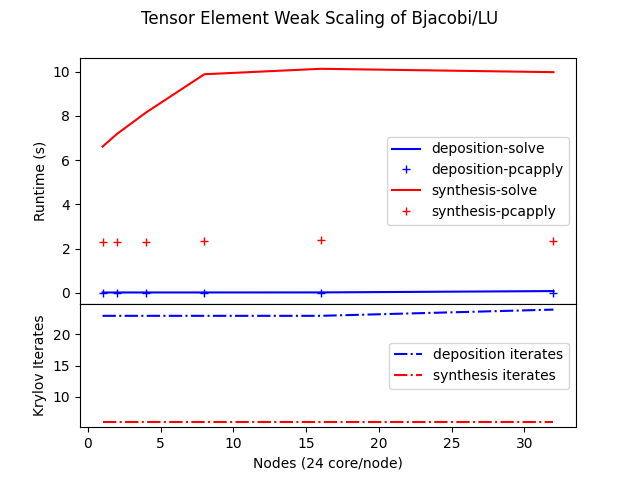}}
    \subfloat[128x128 simplicial cells per node]{\label{fig:simplex_weak_scaling}\includegraphics[width=.45\textwidth]{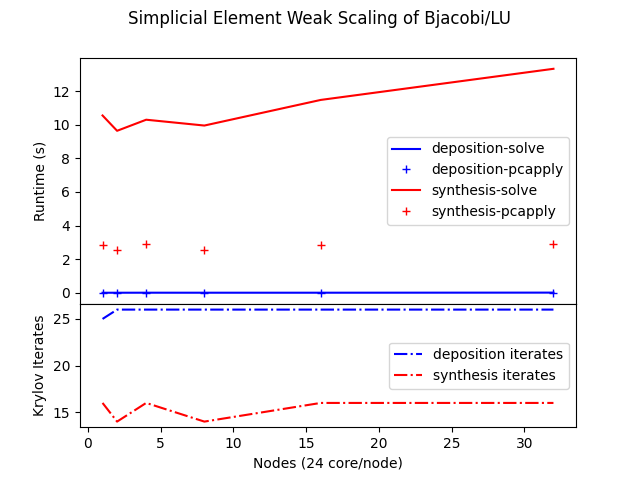}}\\
    \caption{Weak scaling for system solution on tensor and simplicial grids. Both systems have 200 particles per cell and a LSQR/Block Jacobi/LU solver.}
    \label{fig:scaling2}
\end{figure}

 Using LSQR as the outer Krylov solver, with a Block-Jacobi preconditioner built from the normal equations using sparse LU on each block is effective and exhibits good weak scaling. \ref{fig:weak_scaling} displays this behavior for both particle synthesis and deposition on a tensor mesh, and \ref{fig:simplex_weak_scaling} displays results for the simplicial grid.

\begin{figure}[tbhp] 
    \centering
    \subfloat[256x256 tensor cells per node]{\label{fig:checkerboard_weak_scaling}\includegraphics[width=.45\textwidth]{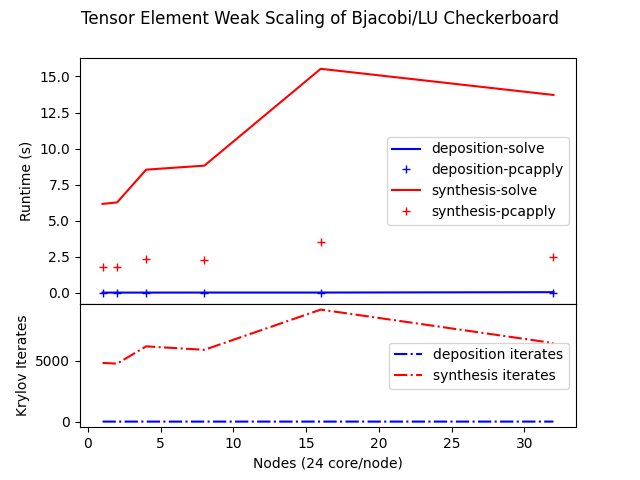}}
    \subfloat[128x128 simplicial cells per node]{\label{fig:checkerboard_simplex_weak_scaling}\includegraphics[width=.45\textwidth]{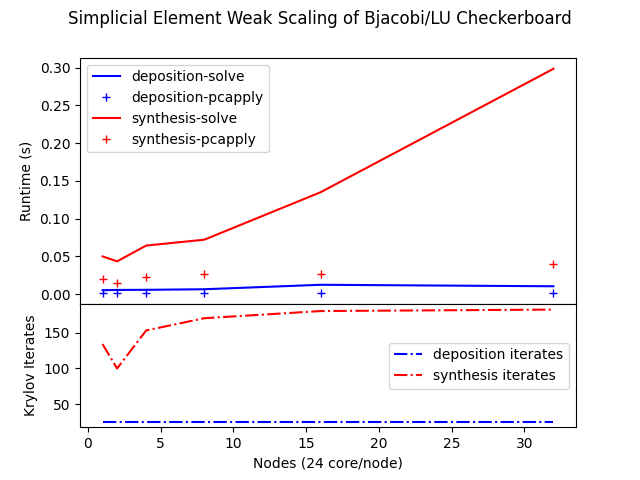}}\\
    \caption{Weak scaling for system solution on tensor and simplicial grids. Both systems are comprised of a "checkerboard" pattern, alternating five and six particles per cell. The LSQR/Block Jacobi/LU solver is used in both cases.}
    \label{fig:checkerboard_scaling}
\end{figure}

Block Jacobi/LU reliably solves the system to within $100*\epsilon_{machine} \sim 2.2*10^{-14}$ regardless of the division of the domain and spacing of the particles when the iterative residual tolerance is low enough, at $\sim\ 10^{-15}$. However, runtimes for moderately large systems can be improved using ASM, as discussed in \ref{sub:asmicc}. Parallel direct solvers such as MUMPS and SuperLU\_dist were considered, but did not decrease the overall time due to increased setup costs, communication overhead, and memory limitations.

In the case of simplicial elements with varying particle number per cell, the number of Krylov iterates scales well, however, at these particle numbers, scatter operations do not scale well and contribute heavily to increases in run time at each increase in the number of nodes and size of the grid. It is also observed in the tensor elements that the increased cost of scatter operations heavily increases run time, with scattering operations on average taking 100x longer in addition to the increase in Krylov iterations necessary to sufficiently solve the system as shown in \ref{fig:checkerboard_scaling}.

\subsection{Additive Schwarz with Incomplete Cholesky}
\label{sub:asmicc}
\begin{figure}[tbhp]
    \centering
    \subfloat[256x256 tensor cells per node]{\label{fig:asmicc_quads_weak_scaling}\includegraphics[width=.5\textwidth]{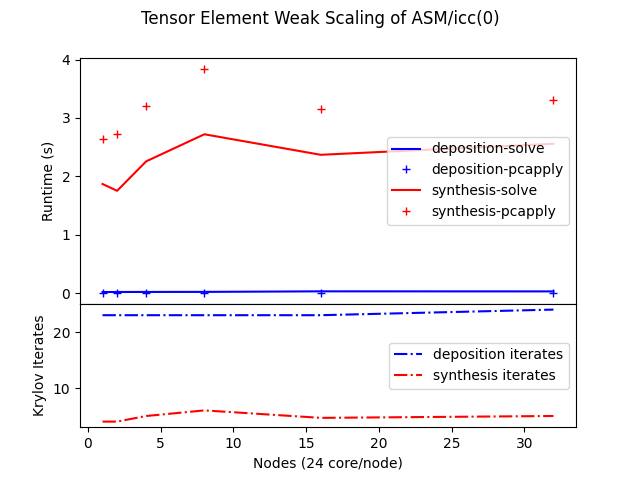}}
    \subfloat[128x128 simplicial cells per node]{\label{fig:asmicc_simplex_weak_scaling}\includegraphics[width=.5\textwidth]{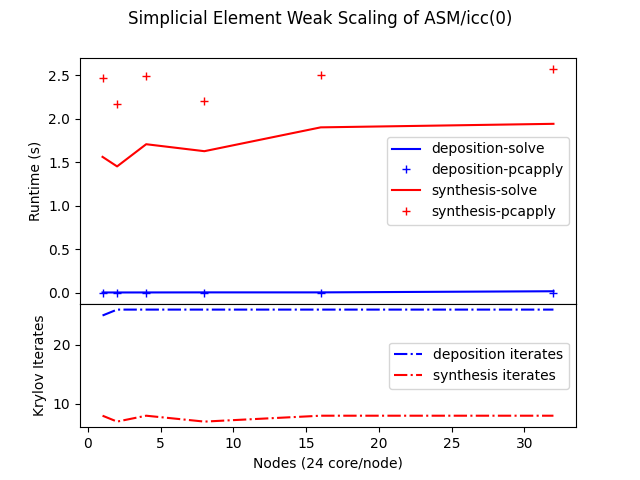}}\\
    \caption{Weak scaling for system solution on tensor and simplicial grids. Both systems have 200 particles per cell and a LSQR/ASM/ICC solver.}
    \label{fig:asmicc_weak_scaling}
\end{figure}

We can accelerate convergence using ASM~\cite{15bcc4ae196041f1a0457f4a82ea9808,b741ebcac0894cee97757442f053cacb} since only weak coupling exists between the blocks and their nearest neighbors. Incomplete Cholesky factorization (ICC)~\cite{Chan1997} is used to precondition the blocks, with the factorization being done in-place. ASM/ICC(0) improves total solve time by a factor of 2 or more in many cases, as seen in~\ref{fig:asmicc_quads_weak_scaling}, and reduced the memory footprint. However, the time due to communication overhead  is noticeable at larger problem sizes, and might be eliminated with a coarse problem. In our largest test case, a run of 1,000,833,800 particles on a $2237 \times 2237$ tensor mesh was converged to machine precision, well beyond what could be achieved with the memory constraints in performing a direct solve.

\begin{figure}[tbhp]
    \centering
    \subfloat[256x256 tensor cells per node]{\label{fig:checkerboard_asmicc_quads_weak_scaling}\includegraphics[width=.5\textwidth]{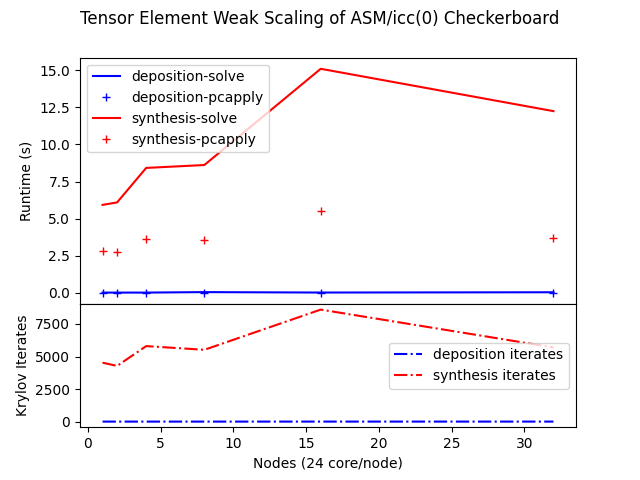}}
    \subfloat[128x128 simplicial cells per node]{\label{fig:checkerboard_asmicc_simplex_weak_scaling}\includegraphics[width=.5\textwidth]{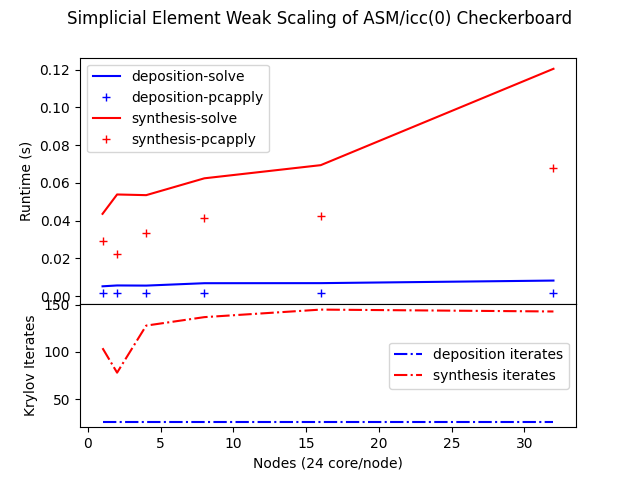}}\\
    \caption{Weak scaling for system solution on tensor and simplicial grids. Both systems are comprised of a "checkerboard" pattern, alternating five and six particles per cell. The LSQR/ASM/ICC solver is used in both cases.}
    \label{fig:checkerboard_asmicc_weak_scaling}
\end{figure}

Compred to the Block Jacobi/Sparse LU, \ref{fig:checkerboard_asmicc_weak_scaling} demonstrates similar behavior of ASM/ICC(0) in regards to scattering operations with low particle number and varying numbers of particles per cell.

An alternative approach to this problem would be to employ a discontinuous Galerkin (DG) discretization for the FEM basis~\cite{KrausHirvijoki2017,HirvijokiKrausBurby2018}. This would completely eliminate coupling between blocks, making the solver purely local. However, this would require a more sophisticated solver for the FEM mass matrix, and also for the associated finite element problem for the continuum physics. For simulations which are already formulated using a DG discretization, this would seem to be a good choice.

\section{Applications}
\label{sec:applications}

Conservative projection operations have applications wherever the PIC discretization is used, but we place particular emphasis on applications in plasma physics codes. For instance, the conservative projection operations we present here have been adapted for, and implemented in, the gyrokinetic PIC code XGC~\cite{KuChangDiamond2009,XGC12018} in order to maintain conservation in their particle to velocity-grid mappings. This code uses the same PETSc solvers detailed above to enforce the algebraic condition in Eq.~\ref{eqn:conservativeProjection}. If conservation is not maintained to high precision, it can cause numerical problems for the simulation, such as artificial heating of the plasma in a steep edge pedestal~\cite{mollen_adams_knepley_hager_chang_2021}. Furthermore, conservative particle-to-grid transformations offer an opportunity to fully leverage the work of Adams and Hirovijoki in the formulation of conservative discretizations of the Landau Collision integral~\cite{AdamsHirvijoki2017} (implemented in PETSc) when used in other PIC codes where conservation at each step is crucial for overall numerical accuracy.

Another potential application is the direct tracking of different material phases without explicit interface tracking. For example, in a hybrid rocket engine, solid paraffin fuel atomizes and is entrained in the liquid oxidizer~\cite{BudzinskiDesjardin2020}. These fuel droplets can be tracked using a particle basis, and interact thermally with the oxidizer using a PIC scheme. Moreover, the transition between mesh based computation of droplet formation and a Lagrangian particle description in the oxidizer flow can be handled using our conservative projection scheme.

\section{Conclusions}
\label{sec:conclusion}
We have derived a method to conservatively project a field between particle and finite element representations, and studied the weak scaling behavior of these solves with different types of preconditioning on the particle mass matrix. LSQR/BJ/LU offers a reliable solve that scales well, but can be limited by the memory footprint. LSQR/ASM/ICC(0) overcomes the memory limitation while maintaining scalability, but can suffer some loss of accuracy near machine precision. This is likely a result of the factor shift introducing some ill-conditioning. This effect is limited and does not always occur, and in general LSQR/ASM/ICC(0) is a good choice for the particle projection solve.

\section*{Acknowledgments} We gratefully acknowledge assistance conducting large scale benchmarks on the Geosolver cluster at the UB Center for Computational Research by Dori Sajdak and Cynthia Cornelius.

\bibliographystyle{siam}
\bibliography{references}
\end{document}